\documentstyle{mn}

\title[On gravitomagnetic precession...]
	{On gravitomagnetic precession around black holes}
\author[A. Merloni et al]
	{ A. Merloni,$^{1}$, M. Vietri$^{2}$,
	 L. Stella$^{3}$\thanks{Affiliated to ICRA} and D. Bini$^{1,4}$
\\$1$ International Center for Relativistic Astrophysics,
                Universit\`a di Roma 1, P.le A. Moro 5 - 00185 Roma, Italy
\\$2$ Universit\'a di Roma 3, Via della Vasca Navale 84 - 00147 Roma, Italy
\\$3$ Osservatorio Astronomico di Roma, Via Frascati, 33 - 00040
                Monte Porzio Catone (Roma), Italy
\\$4$ Istituto per Applicazioni della Matematica,
                 Via P. Castellino 111 - 80131 Napoli, Italy
}

\date{}

\begin{document}

\maketitle

\label{firstpage}

\begin{abstract}
We compute exactly the frequency of Lense--Thirring precession for point
masses in the Kerr metric, for arbitrary black hole mass and specific 
angular momentum. We show that this frequency, for point masses at or close 
to the innermost stable orbit, and for holes with moderate to extreme rotation,
is less than, but comparable to the rotation frequency. Thus, if the 
quasi--periodic oscillations observed in the modulation of the X--ray flux from 
some black holes candidates, BHCs, 
are due to Lense--Thirring precession of orbiting 
material, we predict that a separate, distinct QPO ought to be observed in each 
object. 
\end{abstract}

\begin{keywords}
\end{keywords}

\section{Introduction}

The large effective area, very high time--resolution and excellent telemetry 
of the {\it Rossi} X--ray Timing Explorer (RXTE) have made possible the 
discovery of Quasi Periodic Oscillations (QPOs) in the range 
$\sim 100-1200\; Hz$ 
from a variety of accreting collapsed objects, weakly magnetic neutron stars 
(see van der klis 1998 for a review) and, more surprisingly, 
in black hole candidates 
(BHCs, 
Morgan {\it et al.} 1997; Remillard {\it et al.} 1997).
It has been recently suggested \cite{czc} that these QPOs in BHCs arise through 
Lense--Thirring (LT, 1918) precession of matter from the accretion disks.

Since motion of a point mass in a Kerr metric allows an exact treatment,
and a detailed comparison of Keplerian and Lense--Thirring frequencies 
has not been explicitly carried out 
in the literature up to now, it seems worthwhile to derive 
these quantities for an arbitrary black hole mass and specific angular momentum.
This allows us to clarify the meaning of the precession frequency, about
which some confusion seems to be present in the literature. 
This is the aim of this {\it Letter}. In the last section we shall also
discuss the problems which are raised by this computation, regarding the 
interpretation of Cui {\it et al.} (1998).

\section[]{Bound orbits in the Kerr metric}

In what follows we will consider a test particle of unit mass in motion inside
a Kerr spacetime. The metric, in Boyer-Lindquist coordinates \cite{bl} and
in units $G=c=1$ is
\begin{eqnarray}
\lefteqn{
   ds^2 =  -(1-\frac{2M r}{\rho^2})dt^2
        - \frac{4 aMr\sin^2\theta}{\rho^2}dt d\phi
        + \frac{\rho^2}{\Delta}dr^2} \nonumber \\
&& +\rho^2d\theta^2+
       \frac{\Lambda \sin^2\theta} {\rho^2}d\phi^2 ,
\end{eqnarray}
where
\begin{eqnarray*}
 \rho^2 &=& r^2+a^2 \cos^2\theta  \\
 \Delta&=&r^2+a^2 -2Mr  \\
 \Lambda &=& -\Delta a^2 \sin^2\theta +(r^2+a^2)^2 . 
\end{eqnarray*}
$M$ and $a$ are, respectively, the mass and the specific angular momentum
of the black hole.

As Carter \shortcite{car} first demonstrated, the equation of motion can
be separated, and the resulting equations become: 
\begin{eqnarray}
\rho^2 \dot r &=& \pm \sqrt{R(r)} \label{eqmotr} \\
\rho^2 \dot \theta &=& \pm \sqrt{\Theta(\theta)} \label{eqmotth} \\
\rho^2 \dot \phi &=& (L \sin^{-2}\theta-aE)+a\Delta^{-1}P 
\label{eqmotf} \\
\rho^2 \dot t &=& a(L-aE \sin^2\theta+(r^2+a^2)\Delta^{-1}P 
\label{eqmott} ,
\end{eqnarray}
with
\begin{eqnarray*}
\Theta&=&Q- \cos^2\theta[a^2(1-E^2)+L^2 \sin^{-2}\theta], \\
P&=&E(r^2+a^2)-La \\
R&=&P^2-\Delta[r^2+Q+(L-aE)^2].
\end{eqnarray*}
The dot denotes differentiation with respect to the proper time $\tau$; 
signs in (\ref{eqmotr}) and (\ref{eqmotth}) can be chosen 
independently.
$E$, $L$ and $Q$ are the three constants of the particle motion: $E$ and $L$
are, respectively, the energy and the angular momentum in the azimutal
direction as seen by an observer at rest at infinity; $Q$ is related to 
Carter's constant of motion (see e.g. Chandrasekhar \shortcite{cha} and 
de Felice \shortcite{def}) and characterizes the $\theta$ motion.

As Wilkins \shortcite{wil} showed, bound motion is possible only if
$E^2<1$ and $Q\geq 0$; moreover, for given $Q$ and $L$ and $|E|<1$, there
may be at most one region of binding. Analysis of the $\theta$ effective
potential shows that every orbit either remains in the equatorial plane 
($Q=0$), or crosses it repeteadly ($Q > 0$).
For every bound motion, introducing the angle-action variables, we can
define the three fundamental proper frequencies
\[
1/\tau_{\phi,p}=\nu_{\phi,p}, \; 1/\tau_{\theta,p}=\nu_{\theta,p}, \;
1/\tau_{r,p}=\nu_{r,p},
\]
where $\tau_{\phi,p}$, $\tau_{\theta,p}$ and $\tau_{r,p}$ are the 
proper
time periods for $\phi$, $\theta$ and $r$ motions respectively.
Unlike the Newtonian case of particle motion around a spherically symmetric
central object, where all orbits close and the three fundamental frequencies
are equal, in the Kerr field ($a\neq 0$) there is no degeneracy, i.e.
\[
\nu_{\phi,p}\neq\nu_{\theta,p}\neq \nu_{r,p} .  \] 
The same is also true for coordinate frequencies $\nu_{\phi}$, 
$\nu_{\theta}$ and $\nu_{r}$.

Let us first consider a circular geodesic in the equatorial plane
($\theta=\pi/2$). We have, for the coordinate angular velocities measured by an 
observer static at
infinity \cite{bpt}
\[
\Omega_r=\Omega_\theta=0
\]
\begin{equation}
\Omega_\phi=\frac{2\pi}{\tau_{\phi}}=\frac{d\phi}{dt}=
			\frac{\pm \sqrt{M/r^3}}{1\pm a \sqrt{M/r^3}}; 
\end{equation}
the angular velocity $\Omega_\phi$ deviates from its Keplerian value at small 
radii. The upper sign refers to prograde orbits and the lower to retrograde 
ones. If we slightly perturb a circular orbit introducing velocity components in
the $r$ and $\theta$ directions, we can compute the coordinate frequencies of 
the small amplitude oscillations within the plane (the epicyclic
frequency $\Omega_r$) and in the perpendicular direction (the vertical
frequency $\Omega_\theta$) (Okazaki, Kato \& Fukue \shortcite{okf},
Kato \shortcite{ka}, de Felice \& Usseglio-Tomasset \shortcite{dut},
Perez et al. \shortcite{per}):
\begin{equation}
\Omega_\theta^2=\Omega_\phi^2\biggl[1\mp 
4\frac{aM^{1/2}}{r^{3/2}}+
3\frac{a^2}{r^2}\biggr]
\end{equation}
\begin{equation}
\Omega_r^2=\frac{M(r^2-6Mr\pm 8aM^{1/2}r^{1/2}-3a^2)}{r^2(r^{3/2}\pm
aM^{1/2})^2}.
\end{equation}
In the case of the Schwarzschild metric ($a=0$), there is a partial
degeneracy, as the vertical frequency coincides with the azimutal one. The
epicyclic frequency, instead, is always lower than the other two, reaching
a maximum for $r=8M$ and going to zero at $r=6M$ \cite{okf}. These
qualitative behaviour of the epyciclic frequency is preserved in the Kerr
field ($a\neq 0$), and is a key feature for the existence of trapped
diskoseismic g-modes (Perez et al \shortcite{per}).

\section{Spherical orbits and frame dragging}

We now confine ourselves to the study of those orbits with constant $r$
which are arbitrarily (not infinitesimally) lifted over the equatorial
plane, i.e. with a finite value of $Q$.

The conditions for the stability of a spherical orbit with radius $r=r_0$
are (see eq. (\ref{eqmotr}))
\begin{eqnarray}
R(r_0)= 0 \label{cond:R}\\
\frac{\partial R}{\partial r}|_{r=r_0}=0 \label{cond:dR}\\
\frac{\partial^2R}{\partial r^2} |_{r=r_0}<0 \label{cond:ddR} .
\end{eqnarray} 

Conditions (\ref{cond:R}) and (\ref{cond:dR}) introduce two relations
between $r$ and the constants of motion $E$, $L$ and $Q$, reducing the
free parameters that characterize the orbit to two; thus, given a specific
Kerr black hole (i.e. given the values of $M$ and $a$), a spherical orbit
is completely determinated, for example, by specifying its radius
and the value of $Q$, which fixes the amplitude of motion in the $\theta$
direction (Wilkins \shortcite{wil}).

The motion is open, since the two fundamental frequencies $\nu_\phi$ and
$\nu_\theta$ (proper or coordinate) are incommensurate; then the Fourier
spectra of every function of the position of the test particle will contain
a superposition of the two fundamental frquencies and of all their harmonics
and will be of the kind
\[
\sum_{l=-\infty}^{+\infty}\sum_{m=-\infty}^{+\infty}
C_{lm}e^{i(l\nu_\phi+m\nu_\theta)t+\beta} ,
\]
where $\beta$ is an arbitrary phase.

Therefore, the most natural signals to look for in such a system are the
two fundamental coordinate frequencies themselves and the difference
between them, that, as we will show, coincide
with the unique correct definition of precession frequency of the nodes of
a spherical orbit.

In fact we can compute exactly the coordinate period of the $\theta$ motion:
if we call $\theta_\pm$ (with $\theta_-<\theta_+$) the two roots of the
equation $\Theta(\theta)=0$, we see from (\ref{eqmotth}) that the
particle oscillates on the coordinate sphere between the angles 
$\theta_-$
and $\pi/2+\theta_-$. Dividing (\ref{eqmott}) by (\ref{eqmotth}) and
integrating, we obtain
\begin{eqnarray}
\tau_{\theta}&=& 
4\biggl\{[K(k)-E(k)]\biggl(\frac{z_+}{\beta}\biggr)^{1/2}Ea +
\nonumber \\ &+& \frac{K(k)}{a\sqrt{\beta 
z_+}}\biggl[aL+\frac{P(r^2+a^2)}{\Delta}-
Ea^2\biggr]\biggr\}
\end{eqnarray}
where $\beta=1-E^2$, $k^2=z_-/z_+$ (with 
$z_{\pm}=cos^2\theta_\pm$) and
$K(k)$ and $E(k)$ are the elliptic integrals of the first and second kind,
respectively.

The change of azimuth during one quarter oscillation of latitude is given by
\begin{equation}
\Delta\phi=\frac{1}{a\sqrt{\beta z_+}}\biggl\{L\Pi(-z_-,k)+\biggl[
\frac{a}{\Delta}(2MrE-aL)K(k)\biggr]\biggr\}
\end{equation}
where $\Pi(k)$ is the elliptic integral of the third kind.

An orbit is called co-revolving (or prograde) if $\Delta\phi>0$,
counter-revolving (or retrograde) if $\Delta\phi<0$. If the $\theta$ and
$\phi$ frequencies were the same, $\Delta\phi$ would equal $\pi/2$; it
means that we can define
\begin{equation}
\frac{\nu_\phi}{\nu_\theta}=|\Delta\phi|/\frac{\pi}{2}.
\end{equation}
The angle by which the nodes of a spherical orbit are dragged during each
nodal period is therefore
\begin{equation}
\Delta\Omega=2\pi|\frac{\nu_\phi}{\nu_\theta}-1|
\end{equation}
and, consequently, the coordinate precession frequency of the nodes (or 
{\it Frame-Dragging} frequency) is
\begin{equation}\label{FD}
\nu_{FD}=\frac{\Delta\Omega}{\tau_{\theta}}=|\nu_{\phi}-
\nu_{\theta}|.
\end{equation}
We stress here that this definition is different from the
one given in eq. (2) of Cui, Zhang and Chen \shortcite{czc} ($\nu_{FD}=
\nu_\phi\Delta\Omega/2\pi$), with which it coincides only far from the
source, where $\nu_\phi\simeq\nu_\theta$. But in this case
it would be sufficient to consider the weak-field limit, the well known 
Lense-Thirring equation \cite{lt}, and this point clearly frustrates the
aim of their work.
In fact, approaching the horizon (to which the innermost stable orbit
tends in the limit $a\longrightarrow 1$) their definition leads to a
divergence of the frame dragging frequency, while it can be shown that
\[
\lim_{r\rightarrow r_{hor}}2\pi\nu_{FD}=\Omega_{BH}
\]
where $\Omega_{BH}$ is the angular velocity of the Black Hole \cite{cr,mtw} i.e.
the angular velocity of the Zero Angular Momuntum Observers on the horizon. This
is a relevant difference between this work and that of Cui {\it et al.} (1998).

\begin{figure*}
\vbox to145mm{\vfil 
\caption{The three coordinate frequencies ($\nu_{\phi}$, dashed line, $\nu_{\theta}$, dot-dashed line and $\nu_{FD}$, solid line) of spherical orbits at four diffrent radii as functions of the dimensionless angular moment of an $M=7 M_{\odot}$ black hole.
$r_i$ is the radius of the innermost stable orbit and $r_{peak}$ is the radius of maximum surface emissivity of the disk. $Q$ is always set equal to $1$.}
\vfil}
\label{fig1}
\end{figure*}

We chose to set, in our calculations, $M=7M_{\odot}$, in order to compare
our results with the $300 Hz$ QPO observed from GRO J1655-40 
(Remillard {\it et al.} 1997), the
BHC for which the mass is most accurately measured (Orosz \& Bailyn 1997). We 
considered only direct ({\it i.e.} prograde) orbits. In Figure 1 we plot 
the three frequencies calculated at selected radii as functions of $a$ for 
 $Q=1$. The radii are $r_i$, the radius of the innermost stable circular 
 orbit, which has been calculated solving the quartic equation 
\[
\frac{\partial^2R(r)}{\partial r^2} |_{r=r_i}=0, 
\] 
$r_{peak}=r_i/\eta$ (with $\eta$ slowly varying from $0.62$ to $0.76$ as $a$
goes from $-1$ to $1$), the radius of maximum 
 surface emissivity of the disk (Page \& Thorne 1974), $2r_i$ and $2r_{peak}$. 
The value of $Q=1$ (which corresponds to an `opening angle' of the 
orbit over the equatorial plane which varies from about $3^o$ for $a=0.5$ to 
about $5^o$ for $a=0.99$) was chosen for simplicity, because, exploring the
whole range $Q=0.01-10$, we found the relative changes in the frequencies
raising from $\sim 2\%$ ($a=0.5$) to a maximum of only $\sim 5 \%$ ($a=0.99$).
It is immediately seen that for decreasing radii and increasing values
of $a$ the splitting of $\nu_\phi$ and $\nu_\theta$ increases dramatically.
Correspondingly the frame dragging frequency increases, reaching values that
are comparable to $\nu_\theta$ for $r = 2 r_{peak}$ and $2r_i$ or 
even larger than $\nu_\theta$ for $r = r_{peak}$ and $r_i$.

Figure 2 shows the three frequencies $\nu_{\phi}$,
$\nu_{\theta}$ and $\nu_{FD}$ versus $\nu_{\phi}$ for selected values of the 
angular momentum of the black hole ($a = 0.5,\ 0.9,\ 0.95,\ 0.998$). 
These graphs represent the frequency changes that would take place if, 
for a given black hole, the orbital radius of the precessing matter changed 
(see Section 4).

These results, obtained in a fully general--relativistic framework,
admit a simple interpretation in terms of a Newtonian analogy. 
In the classical gravitational potential due to a spherical star, $\propto
1/r$, the frequencies of motion for a bound orbit in the azimuthal ($\phi$),
radial  and latitudinal ($\theta$) directions are all equal; this
well--known property assures that all orbits close in this potential. Whenever
a small perturbation is introduced, such as that due to the a star's
oblateness, this property is lost and the $\theta$--frequency $\nu_\theta$ 
becomes different from the $\phi$--frequency
$\nu_\phi$. Then the spectrum emitted by a source on this orbit will
contain all harmonics of the type $n\nu_\phi + m\nu_\theta$, with $n,m$ 
integers; of these, the line with, most likely, the largest amplitude is that
at frequency $\nu_\theta-
\nu_\phi$. This, in particular, is the classical precession frequency due to
a Newtonian star not being perfectly spherical. 
Since in classical mechanics departures
from spherical symmetry are always modest, we always find $\nu_\phi -
\nu_\theta \ll \nu_\theta, \nu_\phi$. But in the gravitational field
around a fastly rotating black hole such departures are much more significant,
implying that this inequality is no longer satisfied. In other words,
as departures from a Newtonian potential increase, whether because we are 
moving to a strong--field limit, or because the black hole is rotating
faster, we expect to move toward a situation where $\nu_\phi - \nu_\theta 
\approx \nu_\theta \approx \nu_\phi$. This is exactly what we see happening in 
Figures 1 and 2. 

\begin{figure*}
\vbox to145mm{\vfil 
\caption{The three coordinate frequencies ($\nu_{\phi}$, dashed line, $\nu_{\theta}$, dot-dashed line and $\nu_{FD}$, solid line) of spherical orbits with $Q=1$ for selected values of the angular momentum of the black hole ($a = 0.5,\ 0.9,\ 0.95,\ 0.998$)}
\vfil}
\label{fig2}
\end{figure*}

\section{Application to black hole candidates}

By using the black hole mass and angular momentum which have been measured 
(or indirectly inferred) for several BHCs, Cui {\it et al.} (1998) find a 
reasonably good agreement of the predicted point--mass precession frequencies 
at the radius where the disk emissivity is highest, with the observed QPO 
frequencies. If disk precession is not confined to such a radius
and differential precession takes place at a frequency close to 
the local frame dragging frequency, it remains to be
demonstrated that a sufficiently narrow QPO peak matching the 
observations can be generated as a result of the different precession 
frequencies that might take place at different radii. 
More crucially, it is well--known that in viscous accretion disks 
Lense--Thirring precession of the whole disk is strongly damped (Bardeen
and Petterson 1975, Pringle 1992 and references therein). However there 
appear to be precession modes that are strongly confined to the innermost 
disk regions and only weakly damped (Markovic \& Lamb 1998);  
these are currently being investigated in greater detail. 
The mechanism reponsible for the excitation of these modes remains an open 
question. The perspectives for some kind of resonant excitation driven from 
an azimuthal asymmetry do not appear promising in consideration of the 
black hole "no hair theorem".  

An alternative possibility is that in the innermost disk region
there are individual blobs 
moving like test--particles in the BHC field, executing also LT--precession, 
and modulating the observed X--ray flux either through occultation or because 
they are self--luminous. The existence of discrete blobs is of course not an 
embarrassment for this argument, since their existence is required in all 
scenarios trying to explain QPOs, in particular those involving weakly magnetic 
neutron stars in Low Mass X--ray Binaries (LMXRBs). This tendency of the
disk to form discrete blobs seems to be independent of the nature of the 
accreting source; for instance, Krolik (1998) suggests that it be due to
a need to circumvent the local instabilities of the disk, irrespective 
of the properties of the accreting source. So this tendency might be present
in accretion disks surrounding both neutron stars and black holes.
 
By pushing further the analogy with neutron star LMXRBs, 
where high frequency QPOs are very often observed and succesfully 
interpreted in terms of the Keplerian frequency (note that in LMXRBs
$\nu_{\phi}\simeq\nu_{\theta}$, see below), one would conclude that 
if blobs' precession is responsible for the QPO observed
in BHCs, there is no obvious reason why QPOs reflecting the 
$\phi$ and $\theta$-components of the orbital motion
should not be there.
Indeed, if, according to the model of Cui et al. (1998), the $\sim 300$~Hz QPOs 
of GRS~J1655-40 originate from the frame dragging frequency of blobs off the 
equatorial plane in the innermost disk regions, then
the $\phi$ and $\theta$ frequencies of the orbital motion are
$\nu_\phi \simeq 970$ and $950$~Hz and $\nu_\theta \simeq 670$ and 
$650$~Hz, while $a \simeq0.88$ and $0.96$, respectively in the case 
in which frame dragging QPOs are produced at the innermost stable orbit
or the radius of highest disk emissivity. 
The difference between $\nu_\phi$ and $\nu_\theta$ is large and
two well-separated QPO peaks might be expected. These signals, however, have
not been detected yet. Similar considerations would
apply to the case of the $\sim 67$~Hz QPOs from GRS~1915+105\footnote{The 
harmonic content of the three fundamental frequencies
in the problem at hand will depend on the mechanism responsible for
the generation of the signal(s) (e.g. self-luminous vs. occulting
blobs) and on geometry, and is beyond the scope of this Letter.} 
(Morgan {\it et al.} 1997).

The application of beat frequency models, BFMs, to those neutron stars systems
that show twin kHz QPO peaks, allows us to
identify the higher frequency kHz QPO ($\sim 800-1200$~Hz)
as arising directly from the
Keplerian motion of blobs at the inner egde of the disk
(moreover the neutron star spin frequency is inferred from the
difference frequency of the twin kHz QPOs).
Stella \& Vietri (1998) noticed that the precession frequency of these blobs,
as derived from the neutron star parameters inferred from BFMs, agrees
well with a broad peak around $20-35$~Hz that is apparent in the power
spectra of three sources.
In this model, therefore, $\nu_\phi-\nu_\theta \sim 20-35$~Hz, a separation
that is compable to or smaller than the width of the higher frequency
kHz QPO peak. Therefore it is not surprising that
the signals at $\nu_\phi$ and $\nu_\theta$ are difficult to disentangle 
in the case of neutron star LMXRBs. It should also be noticed that, around
neutron stars with weak magnetic fields, a natural mechanism exists to lift
matter off the equatorial plane, through the interaction with a spinning,
tilted magnetic dipole moment (Vietri and Stella 1998).

In the frame dragging interpretation of BHC QPOs, very specific predictions
are also made in relation to the changes in $\nu_\phi$ and $\nu_\theta$ 
that results from changes in the frame dragging frequency 
$\nu_\phi-\nu_\theta$ (cf. Fig. 2).
This would provide indeed a sensitive diagnostic
to confirm the interpretation and study with unprecedented detail
the motion of matter close to event horizon of a Kerr black hole.

The work of L.S. is partially supported through ASI grants.

\bsp

\label{lastpage}

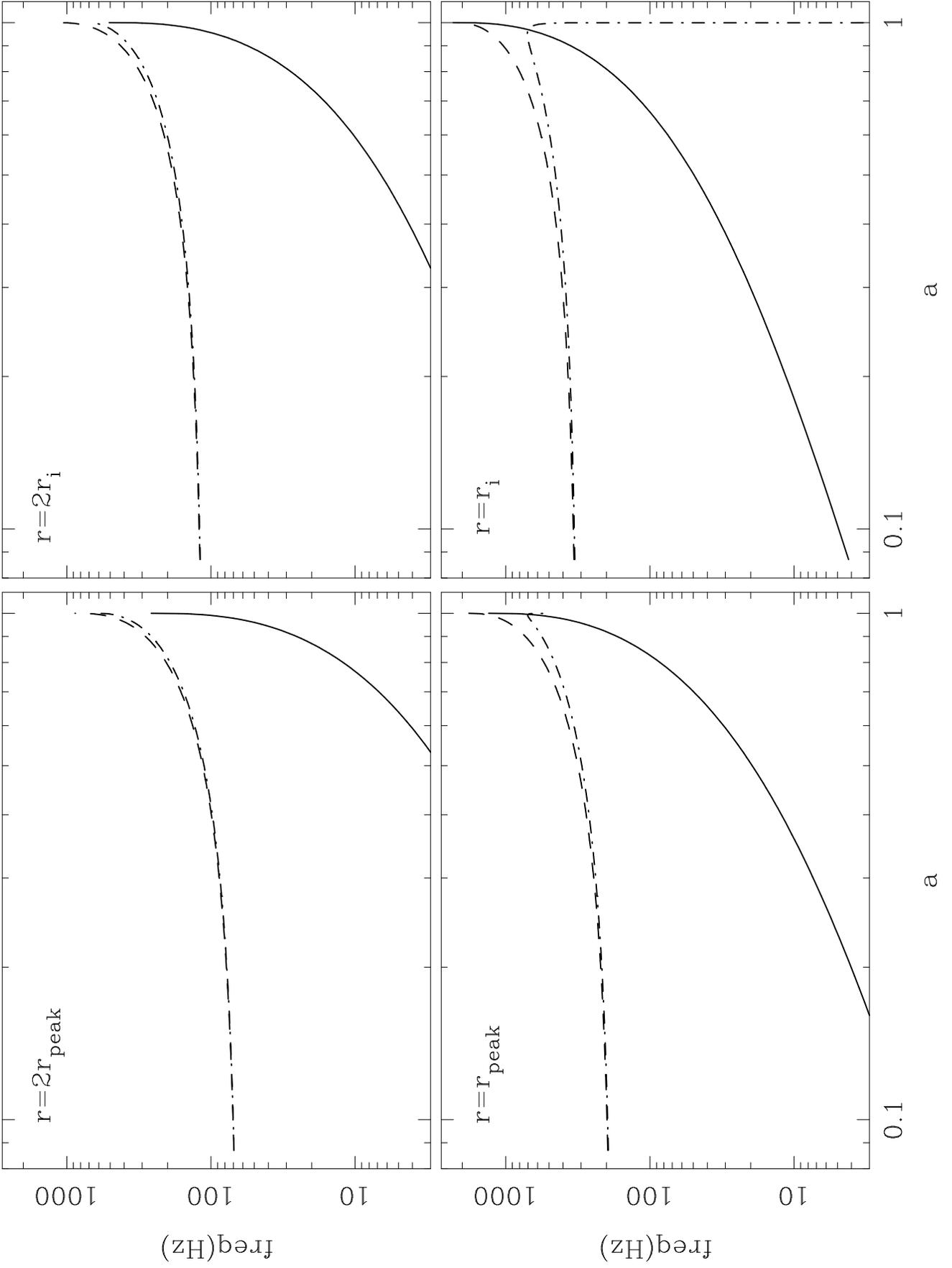
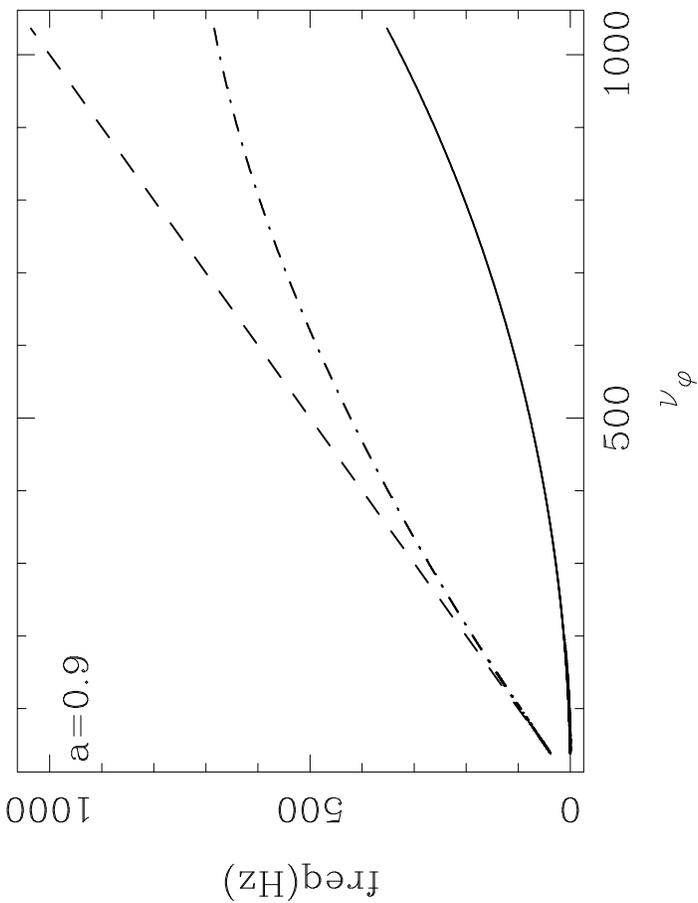
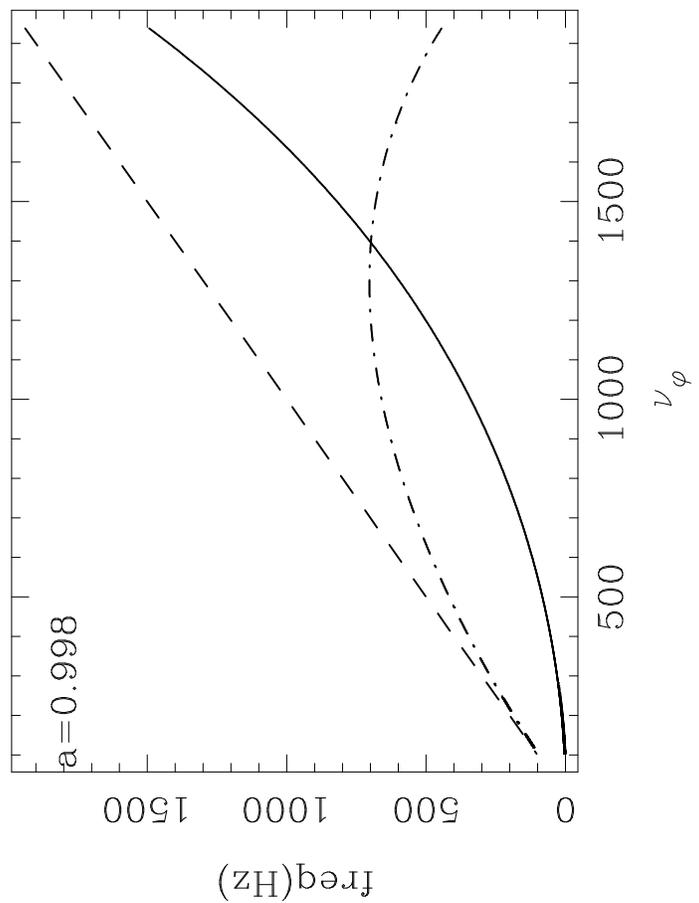
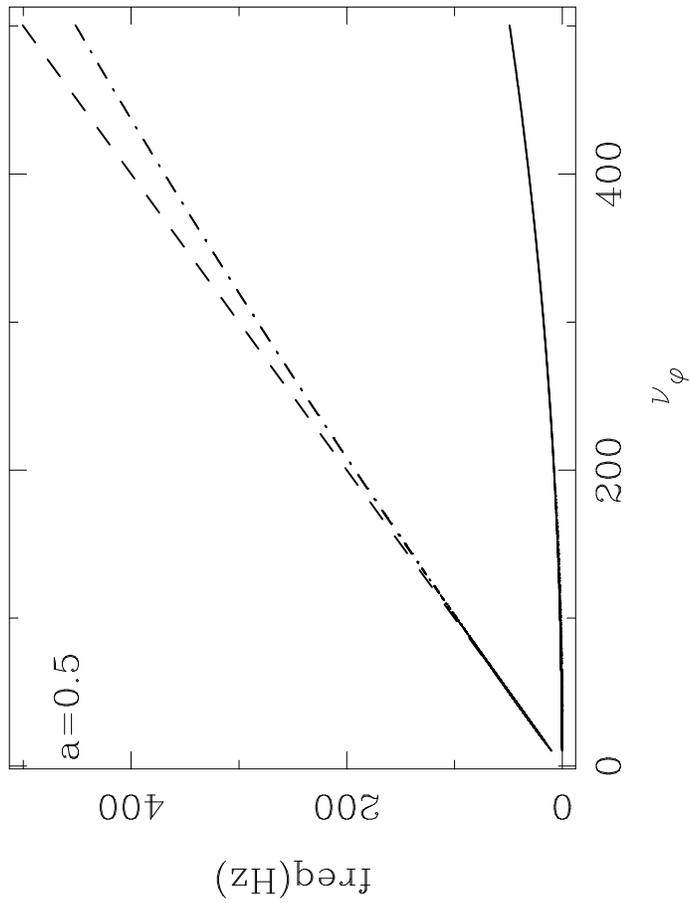
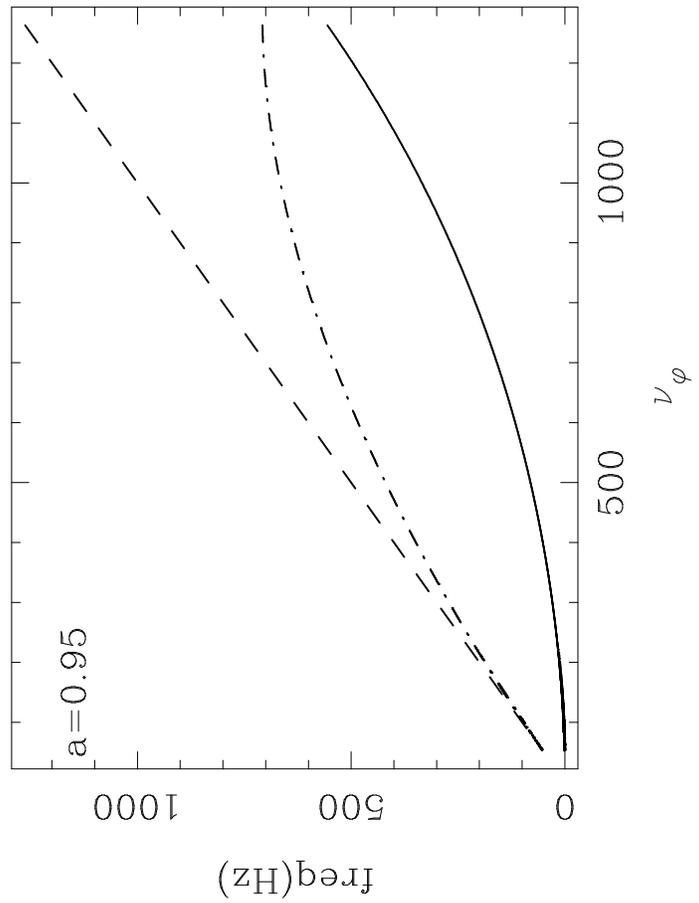
\end{document}